\documentclass[twocolumn,showpacs,preprintnumbers,amsmath,amssymb,prl]{revtex4}

\usepackage{graphicx}
\usepackage{dcolumn}
\usepackage{bm}

\begin{document}

\preprint{Tanabe \textit{et al.}}

\title{Dynamic release of trapped light from an ultrahigh-$Q$ nanocavity via adiabatic frequency tuning}

\author{Takasumi Tanabe}
 \email{takasumi@nttbrl.jp}
\author{Masaya Notomi}
 \email{notomi@nttbrl.jp}
\author{Hideaki Taniyama}
\author{Eiichi Kuramochi}
\affiliation{NTT Basic Research Laboratories, NTT Corporation,\\
3-1, Morinosato Wakamiya Atsugi-shi, Kanagawa 243-0198 Japan}
\affiliation{CREST-JST, 4-1-8, Honmachi, Kawaguchi, Saitama 332-0012, Japan}

\date{\today}

\begin{abstract}
Adiabatic frequency shifting is demonstrated by tuning an ultrahigh-$Q$ photonic crystal nanocavity dynamically.  By resolving the output temporally and spectrally, we showed that the frequency of the light in the cavity follows the cavity resonance shift and remains in a single mode throughout the process. This confirmed unambiguously that the frequency shift results from the adiabatic tuning.  We have employed this process to achieve the dynamic release of a trapped light from an ultrahigh-$Q$ cavity and thus generate a short pulse. This approach provides a simple way of tuning $Q$ dynamically.
\end{abstract}

\pacs{42.70.Qs, 42.79.Nv}

\maketitle

Small high-$Q$ cavities are attractive for various photonics applications that require a strong interaction between light and matter \cite{kippenberg07, yoshie04, englund05, nozaki08, tanabe05apl, tanabe05ol, notomi06prl}.  Of the various micro- and nanocavities, photonic crystal nanocavities (PhC-NCs) have one of the smallest modal volumes \cite{nozaki08, zhang04}.  In particular, the $Q$ of mode-gap confined PhC-NCs has now exceeded a million at a modal volume of $\sim (\lambda /n)^3$ \cite{tanabe07np, noda07}.  This $Q$ value corresponds to a photon lifetime of longer than 1 ns, which presents us with various new ways of controlling light.

Recently, it was predicted that adiabatic shifting of the frequency of light should be possible in such small cavities with a long photon lifetime \cite{notomi06}, where the frequency of trapped photons can be shifted by changing the cavity resonance in a shorter time than the photon lifetime.  This process corresponds to the adiabatic tuning of oscillator systems, which is commonly observed with classical waves such as acoustic waves, (i.e. we can change the tone of a guitar by modulating the tension of the string even after it has been plucked).  This technology may constitute a breakthrough because it allows us to shift the frequency of photons in photonic devices freely, in a way analogous to voltage control in electronic devices.  Shortly after the above theoretical prediction \cite{notomi06}, Preble {\em et al.} \cite{preble07} and McCutcheon {\em et al.} \cite{McCutcheon07} reported on pump-probe experiments to investigate this phenomenon.
They observed modified output spectrum, which suggest that adiabatic wavelength shifting is taking place.  However, the results shown in Ref.~\onlinecite{preble07} always have a peak at the initial frequency that should not exist after the adiabatic tuning.  On the other hand, Ref.~\onlinecite{McCutcheon07} uses much shorter probe pulse, which allows observing single frequency peak.  However, it is still not clear, whether the adiabatic condition is satisfied throughout the process.  In addition, it is not easy to distinguish a shift in the light frequency from a shift in the filtering spectrum of the cavity.
These difficulties arise because they involved comparing time-integrated spectra. Although theoretical support allows interpreting the data \cite{lin08}, it is essential to gain direct view of cavity dynamics to confirm that the frequency shift is actually the result of an adiabatic process.  Therefore, we conducted a series of experiments that include resolving the output from the PhC-NC both temporally and spectrally.

\begin{figure}[b]
 \begin{center}
  \includegraphics[width=3in]{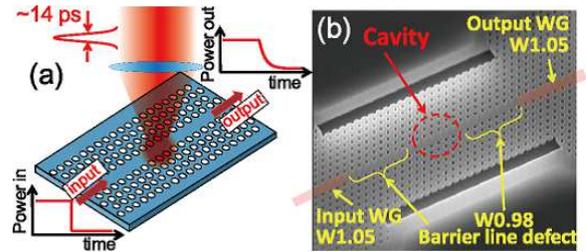}
  \caption{(color online) (a)  Experimental setup.  (b) A scanning electron microscope image of a width-modulated line-defect silicon PhC-NC, where the neighboring air holes are shifted 3, 6, and 9 nm.  The structure of this nanocavity is identical to that described in Ref.~\onlinecite{tanabe07np}.  The PhC-NC is coupled to input and output W1.05 (105\% width of W1) waveguides through W0.98 (98\% width of the W1) line defects.  The length of a W0.98 line defect is $12a$, where $a=420$~nm is the lattice constant.  The W1 waveguide has a width of $a\sqrt{3}$.  The target hole diameter $d$ is 216~nm and the slab thickness $t$ is 204~nm.}
  \label{fig1}
 \end{center}
\end{figure}
The experimental setup for the adiabatic frequency shifting is shown in Fig.~\ref{fig1}(a). The input laser light at the wavelength of the PhC-NC resonance (1608.33 nm) is modulated with a 40-GHz intensity modulator to generate an on-to-off step input waveform (signal).  A pump pulse at a wavelength of 775 nm and an energy of 0.76 pJ is injected into the top of the slab with a spot diameter of 1.3~$\mathrm{\mu m}$ (FWHM) to generate carriers.   About 2.4\% of the injected energy is absorbed by the silicon in accordance with the absorption coefficient, surface reflection, and filling ratio of silicon versus air holes.  The resulting carrier plasma dispersion reduces the refractive index of the silicon.  The timing is set so that the pump pulse is injected immediately after the signal light is turned off.  Recording the decay waveform of the output signal light from the PhC-NC allows us to obtain the photon lifetime directly in the time domain \cite{tanabe07np}.
The PhC-NC we used for the experiment is shown in Fig.~\ref{fig1}(b), which is a cavity that is formed by locally modulating the cut-off frequency of a line defect in a two-dimensional triangular air-hole PhC.  The air holes adjacent to the cavity are shifted a few nanometers away from the line defect to achieve mode-gap confinement \cite{kuramochi06}.  This cavity exhibits an ultrahigh-$Q$ and a small mode volume \cite{tanabe07np, kuramochi06}.  It is in-line coupled to input and output waveguides (W1.05) through W0.98 line defects as confinement barriers.

\begin{figure}[tb]
 \begin{center}
  \includegraphics[width=3.4in]{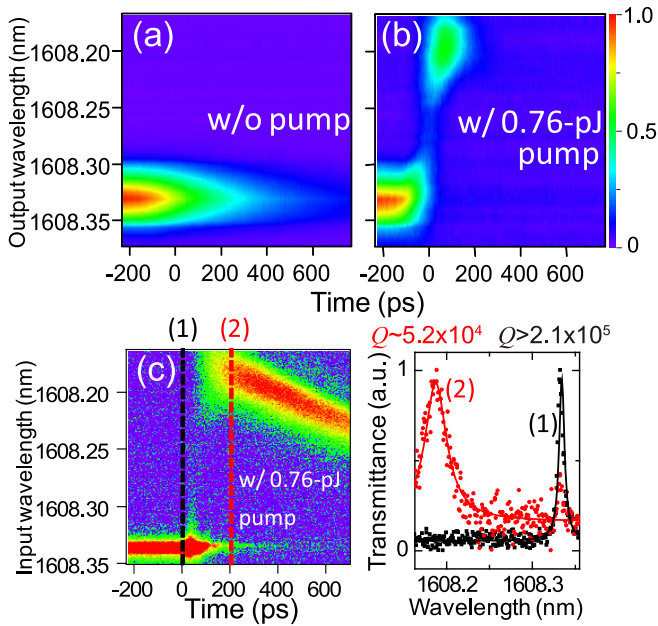}
  \caption{(color online) (a) 2D spectrogram map of time-resolved emission spectra when no pump pulse is applied.  (b) Emission spectra when a 0.76-pJ pulse is applied at 0 ps.  (c)  (left)  Time-resolved transmission spectra, which are the collected transmittance waveforms of the PhC-NC at different input wavelengths.  Carrier lifetime is 2 ns, which is obtained from the slow red-shift after application of the pump.  (right)  Spectrum obtained by taking a cross-sectional image along the wavelength axis of the 2D map at $t=0$ and 200 ps as shown in (1) and (2).}
  \label{fig2}
 \end{center}
\end{figure}
Figure~\ref{fig2}(a) and (b) show spectrograms of the time-resolved wavelength shift of trapped light (hereafter we call this a time-resolved ``emission'' spectrum) released from the PhC-NC.  We used a monochromator as a 0.04-nm bandpass filter and placed it in front of the point at which we measure the output waveform.  The waveform is averaged for 300 times to obtain sufficient signal to noise ratio.  Then we scanned the monochromator wavelength and repeated the same measurement.  Finally we compiled 1D waveforms to obtain a 2D spectrogram that shows the wavelength dynamics of the trapped light in time.
When no pump pulse is injected, the emission spectrum shows a single peak at the wavelength of the input laser light as seen in Fig.~\ref{fig2}(a).  However, when a pump pulse is injected, a new component appears at a shorter wavelength as shown in Fig.~\ref{fig2}(b).  Since only light with a single frequency is present in the cavity, the appearance of a short-wavelength component is direct evidence of a frequency shift. Note that the intensity of this new component decays faster than in Fig.~2(a). This is related to subsequent dynamic $Q$ tuning, which we will discuss later.
More importantly, our time-resolved emission spectra in Fig.~\ref{fig2}(b) confirms that the peak at the initial frequency has shifted to a completely new frequency (i.e. no light remains at the original frequency after the pump), which is a unique characteristic of adiabatic shifting that is not found with other types of conventional nonlinear wavelength conversion.

To investigate this phenomenon in more detail, we measured the time-resolved wavelength shift of the cavity resonance (time-resolved ``transmission'' spectrum) as shown in Fig.~\ref{fig2}(c).
The time-resolved ``transmittance'' spectrum is obtained by scanning the wavelength of the input CW laser light, which is injected into the PhC-NC in 0.8 pm steps.  When a pump pulse ($t=0$~ps) is present, the transmittance of the CW input laser light is modulated as a result of a cavity resonance shift.  By collecting output waveforms for different input laser wavelengths, we obtain a 2D map, which shows the temporally resolved resonance wavelength modulation of the PhC-NC.

Note that Fig.~\ref{fig2}(c) shows the wavelength of the cavity resonance while Fig.~\ref{fig2}(b) shows the wavelength of the trapped light.  A comparison of Fig.~\ref{fig2}(b) and (c) reveals that the light wavelength agrees completely with the shift in the cavity resonance.\cite{note}  This demonstrates that the light frequency shift that we observed in our experiment is actually the result of adiabatic tuning.  Our observations have clarified the distinctive signature of adiabatic frequency shifting for the first time, which became possible because we simultaneously resolved the output both spectrally and temporally.

Next, we show an application of adiabatic wavelength shifting.  We employ this novel optical process to develop on-chip photonic memories and demonstrate photon release with arbitrary timing.  The rapid progress made on photonic technology now allows the all-optical handling of data that were previously carried by electrons.  However, the fundamental differences between photons and electrons mean that achieving an all-optical approach remains a challenge.  One of the advantages of electronics is that electron energy can be controlled with an external field, which provides us with a simple way of manipulating the electric signal flow.  Although there is no straightforward way to change the energy of existing photons, the adiabatic frequency shifting technique will provide an alternative approach for directly manipulating photon energy.  The long photon lifetimes of ultrahigh-$Q$ cavities are clearly attractive as photonic memories, but it is essential to change $Q$ dynamically otherwise there is a trade-off between speed and storage time.  Various techniques have been reported for changing $Q$ dynamically \cite{xu07, tanaka07, notomi07}, by employing sophisticated interference methods, but we here describe a method that does not require interference control; a method based on adiabatic frequency tuning.

\begin{figure}[tb]
 \begin{center}
  \includegraphics[width=3.5in]{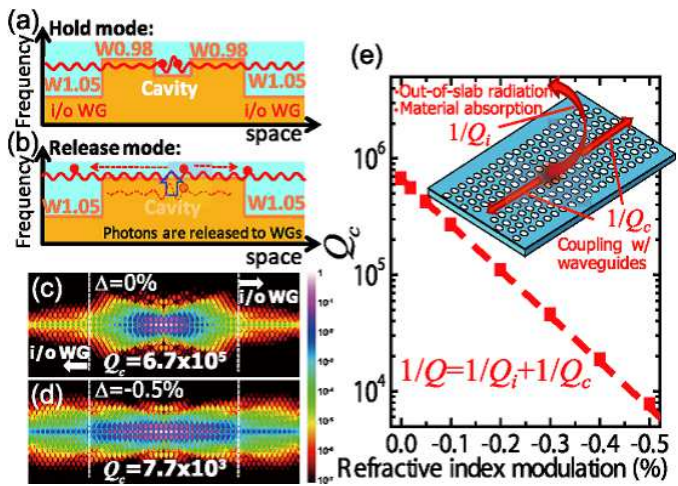}
  \caption{(color online) (a)  Operating principle of the dynamic tuning of $Q$.  Schematic frequency band diagram in the spatial coordinates in the high $Q$ mode.  The light blue area is the frequency regime where propagation is allowed.  The yellow area is the mode-gap region of a line defect.  (b) Low-$Q$ mode.  (c) Mode profile calculated with the 3D finite difference time domain method for a PhC-NC coupled to the input and output waveguides.  The structure is the same as that in Fig.~\ref{fig1}(b) except that $d=0.48a$.  WG:  waveguide. (d)  The refractive index at the cavity is modulated with a Gaussian distribution (FWHM:  $6a$).  (e) Refractive index modulation versus $Q_c$.  The inset shows the definitions of $Q_c$ and $Q_i$. $Q_i$ does not change significantly within this range of modulation.}
  \label{fig3}
 \end{center}
\end{figure}
Now, we recall the principle of light confinement with this PhC device.  Figure~\ref{fig3}(a) is a schematic diagram of a mode-gap confined PhC-NC.
Photons are initially trapped in the cavity by W0.98 confinement barriers.  When we change the refractive index of the cavity within the photon lifetime, the frequency of the trapped light is adiabatically shifted in proportion to the shift in the cavity resonance, as shown in Fig.~\ref{fig2}(b).  At this new frequency, the confinement barriers are shallower than in their initial condition, and the trapped photons are coupled more strongly with the waveguides (Fig.~\ref{fig3}(b)).  Thus the trapped photons are released as a pulse that is much shorter than the original photon lifetime.  The key feature of this cavity is that strong light confinement is achieved by a small structural modulation, which should enable us to change $Q_c$ ($Q_c$ represents the cavity-waveguide coupling) with a small refractive index modulation.

This method of releasing photons dynamically from an optical cavity is analogous to the operation of a charge transfer device employing an field-effect transistor (FET) \cite{fujiwara08}, as the trapped electrons (photons) are released when an applied gate voltage (pump pulse) shifts the electron energy (photon frequency) above one of the potential (mode-gap) barriers.  This type of energy shifting is widely used in electronics devices, so it has the potential to expand the possible applications of photonic technologies if it becomes available in the photonics field.

First, we undertook a numerical study to determine whether the above mechanism works with the available small index change.  The $Q$ of a cavity is given by $Q^{-1}=Q_i^{-1}+Q_c^{-1}$, where $Q_i$ is the inverse of the intrinsic cavity loss (radiation and absorption) (Fig.~\ref{fig3}(e) inset).  Without index tuning, the cavity exhibits a $Q_i$ of $1.2\times 10^8$ and a $Q_c$ of $6.7\times 10^5$ (Fig.~\ref{fig3}(c)).  Reducing the refractive index of the cavity causes $Q_c$ to decrease because the cavity mode couples more strongly with the waveguides, as shown in Fig.~\ref{fig3}(d).  Figure~\ref{fig3}(e) shows the way in which $Q_c$ is related to the refractive index modulation.  Although, complicated configurations were previously required \cite{xu07, tanaka07} we achieved the sensitive tuning of $Q_c$ with a very simple configuration.

\begin{figure}[tb]
 \begin{center}
  \includegraphics[width=2.2in]{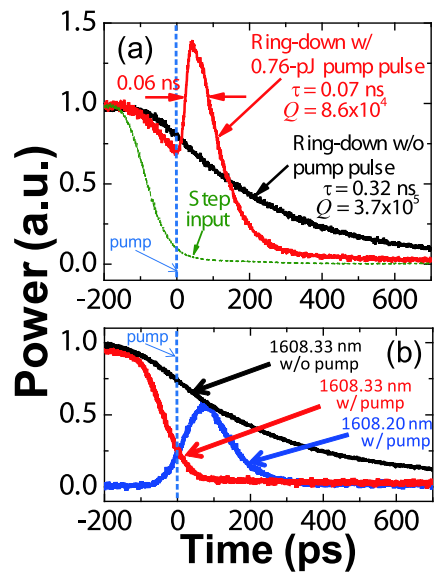}
  \caption{(color online) (a) On-to-off step input (dotted line).  Recorded output from the PhC waveguide, which is the light coupled from the PhC-NC, with and without a pump pulse.  The refractive index is modulated by $\sim 0.01$\%.  The exponential fit after the pump is 0.07 ns, which corresponds to a $Q$ of $8.6 \times 10^4$.  (b) Spectrally resolved waveforms obtained by taking line plots along the temporal axis of Figs.~\ref{fig2}(a) and (b).}
  \label{fig4}
 \end{center}
\end{figure}
Next, we undertook an experimental demonstration of short pulse generation. As readers may be aware, the required operation is essentially the same as that shown in Fig.~\ref{fig2}.  That is, we should observe short pulse generation from the output waveguide under the same experimental conditions but without using the monochromator. Fig.~\ref{fig4}(a) is the recorded output signal.  When no pump pulse is injected, we obtain a photon lifetime of 0.32 ns, which corresponds to a $Q$ of $3.7\times 10^5$.  From the transmittance, we obtain $Q_c \simeq 9.4 \times 10^5$ and $Q_i \simeq 6.2 \times 10^5$.  When we inject a pump pulse, we find that the output power increases significantly, and a short pulse is extracted from the cavity with a width of $\sim 0.06$ ns, which is much shorter than the original photon lifetime.  The instantaneous power from the nanocavity immediately after the pump increased about 2 times.  Therefore, we know that $Q_c$ is dynamically tuned and is reduced by a factor of 2 in this experiment.  Note that the power at the waveguide does not increase unless $Q_c$ is changed.  In fact, we observed no power enhancement when we changed $Q_i$ \cite{tanabe07np}.  This means that Fig.~\ref{fig4}(a) depicts a direct demonstration of the dynamic tuning of $Q_c$, which releases a short pulse from an ultrahigh-$Q$ PhC-NC into the waveguide.  The direct demonstration of the release of a short pulse is exactly what we need for photonic memory operation.  We should emphasize that this operation is achieved with only a simple frequency-tuning process.

Here, we comment on the extraction efficiency $\eta$ \cite{eta}.
It should be noted that $\eta$ is greater than 100\% when the operation is ideal.  If we can reduce $Q_c$ without changing $Q_i$, we can redirect the flow of the light that was initially coupled to the out-of-slab radiation into the waveguide coupling mode, resulting in an increase in the total energy at the output waveguide.  Since $\eta$ is proportional to the $Q_c$ ratio before and after the modulation, we can obtain $\eta =200$\% when $Q_c$ is changed by a factor of 2.  However $\eta \simeq 80$\% in Fig.~\ref{fig4}(a).  This is due to an increase in the free-carrier absorption, which reduces the $Q_i$ of the cavity.  In Fig.~\ref{fig4}(a), $Q_i$ is $\sim 1.1\times 10^5$ after the pump, which is obtained from the slope of the decay and $Q_c$ after the pump.  Note that at a cost of $\eta$, we can obtain an even shorter output pulse (0.01~ns at 1.46-pJ pump).

Finally, we investigate the frequency characteristics of Fig.~\ref{fig4}(a), because the frequency tuning process obtained in Fig.~\ref{fig2}(b) is the key to this operation.  As part of this investigation, we plot the waveforms for the original wavelength and the shifted wavelength in Fig.~\ref{fig4}(b) by employing a cross-sectional image taken from along the temporal axis of the spectrograms in Figs.~\ref{fig2}(a) and (b).  When there is no pump pulse, the output exhibits a smooth exponential decay.  However when a pump pulse is injected, the output waveform for the original wavelength exhibits a sharp decay.  Furthermore, a shorter wavelength component starts to appear.  A comparison of this result with Fig.~\ref{fig4}(a), which shows the dynamic tuning of $Q_c$, reveals complete consistency.  So, we find that the short pulse that is read out from the cavity actually has a shorter wavelength than the original laser light.  The combination of this result with Fig.~\ref{fig2} unambiguously confirms that the operation in Fig.~\ref{fig4}(a) is indeed driven by adiabatic frequency shifting.

In this study, we resolved the output from a dynamic ultrahigh-$Q$ nanocavity both spectrally and temporally and showed directly that adiabatic frequency shifting is present.  In addition, we demonstrated short pulse release from an ultrahigh-$Q$ nanocavity using adiabatic frequency shifting whose operation is analogous to that of FET based devices. This demonstration shows that adiabatic frequency shifting may expand the possibilities of controlling light in a chip beyond the capacity of conventional photonic technologies.  It should be noted that this classical frequency shifting does not involve multi-photon processes, and thus the conversion efficiency and frequency shift are independent of the number of photons.  This allows the approach to work at a single photon level, which even makes this technology attractive for developing quantum information processing on chip.


\begin{thebibliography}{99}
\bibitem{note}Light propagates with a delay of about twice the photon lifetime ($\tau_{\mathrm{ph}}$) through a cavity \cite{tanabe07np}.  Since the modulated $\tau_{\mathrm{ph}}$ is 0.07~ns [Fig.~\ref{fig4}(a)], short wavelength light transmits with a delay of $2\tau_{\mathrm{ph}}=0.14$~ns, as shown in Fig.~\ref{fig2}(c).  On the other hand, the trapped light is released to the waveguide at a speed of $\tau_{\mathrm{ph}}=0.07$~ns as agrees with the appearance of short wavelength component in Fig.~\ref{fig2}(b).
\bibitem{eta}
	 We define $\eta$ as,
$\eta = \frac{\int^\infty_{t_{\mathrm{pump}}} P_\mathrm{w/}(t) dt}{\int^\infty_{t_{\mathrm{pump}}}P_\mathrm{w/o}(t) dt}\times 100.$
$P_\mathrm{w/}$ and $P_\mathrm{w/o}$ are recorded waveforms from the PhC-NC at the output PhC waveguide with and without a pump pulse, respectively.  $t_\mathrm{pump}$ is the timing when a pump pulse is applied ($t_\mathrm{pump}=0$~ps in Fig.~\ref{fig4}(a)).
\end{thebibliography}
\end{document}